\documentclass[preprint,12pt,amsmath,amssymb,nofootinbib]{revtex4}
\usepackage{graphicx} % Include figure files

\newcommand{\smQ}{{\rm\scriptscriptstyle Q}}
\newcommand{\smL}{{\rm\scriptscriptstyle L}}
\newcommand{\smC}{{\rm\scriptscriptstyle C}}
\newcommand{\smS}{{\rm\scriptscriptstyle S}}
\newcommand{\smR}{{\rm\scriptscriptstyle R}}
\newcommand{\smI}{{\rm\scriptscriptstyle I}}
\newcommand{\smH}{{\rm\scriptscriptstyle H}}
\newcommand{\smT}{{\rm\scriptscriptstyle T}}
\newcommand{\smzero}{{\rm\scriptscriptstyle 0}}
\newcommand{\smQM}{{\rm\scriptscriptstyle QM}}
\newcommand{\smFD}{{\rm\scriptscriptstyle FD}}
\newcommand{\smBPS}{{\rm\scriptscriptstyle BPS}}
\newcommand{\smDelQ}{{\rm\scriptscriptstyle \Delta Q}}
\newcommand{\smDelC}{{\rm\scriptscriptstyle \Delta C}}

\begin{document}

\preprint{\hfill 
LA-UR-08-2826 ~~~Phys. Rev. E {\bf 79} (2009) 066407}

\title{
    Temperature equilibration in a fully ionized
    plasma: \\
    electron-ion mass ratio effects
}

\author{Lowell S. Brown and Robert L. Singleton Jr.}

\affiliation{
     Los Alamos National Laboratory,
     Los Alamos, New Mexico 87545, USA
}

\date{17 February 2009}

\begin{abstract}
  Brown, Preston, and Singleton (BPS) produced an analytic calculation 
  for energy exchange processes for a weakly to moderately coupled plasma: 
  the electron-ion temperature equilibration rate and the charged 
  particle stopping power. These precise calculations are accurate to 
  leading and next-to-leading order in the plasma coupling parameter, 
  and to all orders for two-body quantum scattering within the
  plasma. Classical molecular dynamics  can provide another approach 
  that can be rigorously implemented. It is therefore useful to compare
  the predictions from these two methods, particularly since the
  former is theoretically based and the latter numerically. An
  agreement would provide both confidence in our theoretical machinery
  and in the reliability of the computer simulations. The comparisons 
  can be made cleanly in the purely classical regime, thereby avoiding 
  the arbitrariness  associated with constructing effective potentials 
  to mock up  quantum effects. We present here the classical limit of 
  the general result for the temperature equilibration rate presented 
  in BPS. In particular, we examine the validity of the 
  $m_{\rm electron}/m_{\rm ion} \to 0 $ 
  limit used in BPS to obtain a very simple analytic evaluation of the 
  long-distance, collective effects in the background plasma. 
    
\end{abstract}

\maketitle

Brown, Preston, and Singleton~\cite{bps} (BPS) have performed a
controlled first principles analytic calculation of Coulomb energy
exchange processes for weakly to moderately coupled plasmas, namely
the electron-ion temperature equilibration rate and the charged
particle stopping power. These calculations are accurate to leading
and next-to-leading order in the plasma coupling parameter $g\sim (
e^2 / \lambda_{\rm Debye}) / T $, and to all orders in the quantum
parameter $\bar\eta \sim e^2 / \hbar \bar v_\smT$ for the two-body
quantum scattering within the plasma, where $\bar v_T$ is a thermal
averaged electron velocity.  Here we shall examine the electron-ion
temperature equilibration rate in some detail, particularly with
regard to recent molecular dynamics calculations \cite{HM80s,dd}.

In general, we write the energy density exchange rate between the
electrons and ions in a plasma as\,\footnote{ 
  We use the rate of change of electron-ion energy density $d{\cal
  E}/dt$ on the left hand-side of Eq.~(\ref{dedteeI}), rather than the
  corresponding temperatures $dT/dt$, because, in general, the
  conversion between ${\cal E}$ and $T$ entails higher order plasma
  corrections in the equation of state.  However, as noted in
  Ref.~\cite{bps}, a clean separation into separate electron and ion
  energy components within the plasma can be made only up to the order
  $g$ to which we work here. The kinetic energy does not depend upon
  the coupling $g$, and its time derivative is of order $g^2 \, \ln
  g$. On the other hand, the potential energy starts at order $g^2$,
  and its time derivative is of higher order than we are
  calculating. At higher orders, collective, plasma potential energies
  come into play, and this clean separation into electron and ion
  energies cannot be made.
}
\begin{eqnarray}
  \frac{d{\cal E}_{e \, \smI}}{dt} 
  =   
  -\,{\cal C}_{e \,\smI}\left(T_e - T_\smI \right) \,,
\label{dedteeI}
\end{eqnarray}
where the rate parameter that appears here, 
\begin{equation}
  {\cal C}_{e\smI}
  = 
  \frac{\kappa_e^2}{2\pi} \, \left(\frac{m_e}{2\pi\, T_e}\right)^{1/2} 
  \omega_\smI^2 \, \ln\Lambda \,,
\label{rate}
\end{equation}
serves to define the ``Coulomb logarithm'' $ \ln\Lambda $ for
this specific process. The electron Debye wave number $\kappa_e = 1 /
\lambda^{\rm electron}_{\rm \, Debye}$, and
\begin{equation}
  \omega_\smI^2 = {\sum}_i \, \omega_i^2 
\end{equation}
is sum of the squares of the ion plasma frequencies. Throughout this
paper we shall measure temperature in energy units.

For plasma conditions that often occur in ICF capsules, the coupling
$g$ is weak to moderate and the electron temperature $T_e$ is much greater
than the binding energy $\epsilon_\smH$ of the hydrogen atom, $ T_{e}
\gg \epsilon_\smH $. Since $ \bar v_\smT \sim \sqrt{T_e / m_e}$, the
quantum parameter can be alternatively be written as $\eta \sim
\sqrt{\epsilon_\smH/T_e}$. Thus, for these ICF conditions, the quantum
parameter is small, $\eta \ll 1 $, which corresponds formally to
$\hbar$ being large, and so the scattering must be described quantum
mechanically. The electron mass $m_e$ is much smaller than an ion mass
$m_\smI$ in the plasma. Taking advantage of $m_e/m_\smI \lll 1$, BPS
made use of a sum rule for the contributions of collective
long-distance effects in the $m_e \to 0$ limit and were able to find
a simple expression for the Coulomb logarithm under these conditions:
\begin{eqnarray}
  \ln\Lambda_\smBPS^\smQM
  =
  \frac{1}{2}\left[\ln\!\left\{\frac{8 T_e^2}{\hbar^2 \omega_e^2}
  \right\} - \gamma - 1 \right] \,,
\label{bpsrateQM}
\end{eqnarray}
where $\gamma=0.57721 \cdots$ is the Euler constant, and $\omega_e$ is
the electron plasma frequency.  The result (\ref{bpsrateQM})
corresponds to Eq's.~(3.61) and (12.12) of BPS~\cite{bps}, with a
small transcription error corrected when BPS passed from their
Eq.~(12.43) to Eq.~(12.44).

Molecular dynamic (MD) methods\,\cite{HM80s} have been used to extend
such perturbative results into non-perturbative regimes. However, most
MD methods employ ad-hoc potentials that mock up quantum-mechanical
effects in a fashion that has little theoretical basis. Consequently,
even in the perturbative regime, it is problematic to compare such MD
results with the result of a systematic calculation such as that
given in Eq.~(\ref{bpsrateQM}).  To surmount this problem, Dimonte and
Daligault\,\cite{dd} (DD) have recently performed purely
classical~\footnote{
  Dimonte and Daligault have obtained unambiguous and accurate results
  for a well-defined problem. They do not treat quantum-mechanical
  effects, and thus their work is not directly applicable for the
  conditions that appear, for example, in an ICF capsule.
} 
MD simulations with the Coulomb potential. The MD simulations of DD
give results that can be directly compared with the classical limit of
the BPS calculation. As with the extreme quantum limit
(\ref{bpsrateQM}), for light electrons the classical result also takes
a simple form,
\begin{eqnarray}
  \ln \Lambda^\smC_\smBPS
  =
  {\sum}_i \, \frac{\omega_i^2}{\omega_\smI^2} \,
  \ln\!\left\{ 
  \frac{4 T_e} {|e e_i| \, \kappa_e} \right\}
  - 2 \gamma\, - \frac{1}{2} \,.
\label{mezeroLC}
\end{eqnarray}
This expression follows from adding terms (12.25) and (12.44) of BPS,
except for the trivial modification to ordinary Gaussian units from
the rationalized electrostatic units employed by Ref.~\cite{bps}. 
Here, $e$ and $e_i=Z_i\, e$ are the values of the electron and ion
charges.  For the electron-proton plasma considered by DD, the
classical BPS result can be written as $\ln \Lambda^\smC_\smBPS =
\ln\{C/g_e\}$, where $g_e = e^2 \kappa_e/T_e$ is the plasma coupling,
and $C=4 e^{-(2\gamma + 1/2)}=0.7648 \cdots$~. For small values of
$g_e$, the regime in which the BPS formalism is valid, DD have found
agreement with expression (\ref{mezeroLC}) to within their statistical
error of 5\,\%.

To avoid the formation of unstable configurations of two particles of
opposite charge that experience an attractive force, DD work with
electrons and ions of the same charge moving in an implicit
neutralizing background charge density of opposite sign.  Since it is
either the square or the absolute value of the electron and ion
charges that enter into the perturbative regime, as made explicit by
the expressions (\ref{rate}) and (\ref{mezeroLC}), this sign change
causes no problem for small coupling~\footnote{
  However, in the non-perturbative regime where the plasma coupling
  $g$ becomes of order one or larger, the relative sign of the
  electron and ion charges is important.  For example, the first
  correction to the perturbative results quoted in the text 
  has an overall coefficient involving $(e e_i)^3$ that is 
  negative in the physical case. This is experimentally confirmed
  in the differing ranges of positive and negative pi mesons
  stopping in nuclear emulsions \cite{bark}.
}. 
To increase the speed of convergence of the simulations, DD use an
`electron' mass $m_e$ and a single species of `ion' mass $m_\smI$
whose ratio is on the order of 1/100 rather than the physical value of
1/1836 for the electron and proton. This larger choice of electron-ion
mass ratio allows them to obtain accurate numerical results with high
statistics, and it in no way alters the relevance of their work in
testing analytic results for which this mass ratio can be changed at
will. However, to properly compare with BPS in this mass regime, one
must now include finite electron mass corrections to
Eq.~(\ref{mezeroLC}). Some are trivial kinematic corrections.  Others,
which we shall denote by $\Delta$, arise from long distance collective
effects and are non-trivial. For the parameter regime considered in DD
\cite{dd}, we shall find that these corrections are less than their
5\% statistical error.

Before turning to the calculation of the electron mass corrections for
the classical case, we note that the extreme quantum
limit~(\ref{bpsrateQM}) contains exactly the same non-trivial long
distance correction $\Delta$, in addition to other trivial kinematic
corrections. Therefore, the calculation in the text is also relevant
for the extreme quantum case. For plasmas in the ICF regime, these
corrections may be of comparable size to the degeneracy 
corrections~\footnote{ 
  Section III of Ref.~\cite{bs} contains a detailed pedagogical
  account of the method of dimensional continuation that we employ.
  Section IV of this paper reviews the ad-hoc nature of the
  ``Convergent Kinetic Equations'' that have been used in the
  literature. The work of Kihara and Aono\,\cite{KA} which examines
  the energy relaxation of a slow particle in a plasma falls into this
  latter category.
} 
calculated in Ref.~\cite{bs}. To have a place in which all
the small corrections to the quantum limit (\ref{bpsrateQM}) are
collected, we quote results of Ref.~\cite{bs} in the Appendix.  These
entail not only the first Fermi-Dirac correction when the electron
density starts to become large, but also the first classical
correction when the quantum parameter $\eta$ starts to become
large. Thus, we shall have a convenient reference to all of our
results on the energy relaxation rate relevant to ICF recorded in the
Appendix.

To begin our development, we note that the work of BPS \cite{bps}
expresses the coefficient appearing in Eq.~(\ref{dedteeI}) as the sum
of three contributions,
\begin{eqnarray}
  {\cal C}_{e \, \smI}
  =
  \Big(  {\cal C}^\smC_{e \, \smI \,,\, \smS} +
   {\cal C}^\smC_{e\, \smI \,,\,\smR} \Big)
  +
  {\cal C}^{\smDelQ}_{e \, \smI} \,,
\label{Cabclqm}
\end{eqnarray}
with the three terms given by Eq's~(12.25), (12.31), and (12.50) in
BPS. The first term ${\cal C}^\smC_{e \, \smI \,,\, \smS}$ is the
short-distance classical contribution.  The second term ${\cal
C}^\smC_{e \, \smI \,,\, \smR}$ is the contribution from
long-distance, cooperative, dielectric corrections.  These two terms
comprise the complete classical rate coefficient, which is the main
topic of this note. The remaining term ${\cal C}^\smDelQ_{e \, \smI}$
is the complete quantum correction (which vanishes in the formal limit
$\hbar \to 0$).  The method used by BPS enables these terms to be
unambiguously and precisely calculated to leading and first subleading
order in the plasma coupling --- expressed in terms of the plasma
number density $n$, this is a unique evaluation to the formal orders
$n \ln n$ and $n$.  Since the expression for ${\cal C}^\smDelQ_{e \,
\smI}$ is not needed for the discussion in the text, and since its
ingredients are somewhat complex, we shall relegate it to the
Appendix.  The quantum result is, however, an essential contribution
in practical ICF applications.

The first term in Eq.~(\ref{Cabclqm}), the short-distance classical
scattering contribution, reads~\footnote{
  This is Eq.~(12.25) of Ref.~\cite{bps} in somewhat different
  notation, and with the arbitrary wave number $K$ set to the electron
  Debye wave number $\kappa_e$ for convenience; see also Eq.~(B8) of
  Ref.~\cite{bs}.
}
\begin{equation}
  {\cal C}^\smC_{e\smI,\smS} 
  = 
  -{\sum}_i \kappa_e^2 \, \kappa_i^2 \, 
  \frac{ (T_\smI^2 m_e T_e^2 m_i)^{1/2}}{\left( T_\smI m_e + 
  T_e m_i \right)^{3/2} } \,\left( \frac{1}{2\pi} \right)^{3/2}\, 
  \left[\ln\!\left\{\frac{ Z_i \, g_e \, T_e}{4\,m_{ei}\,V^2_{ei}}
  \right\} +  2 \gamma\,  \right] \,. 
\label{classicdone}   
\end{equation}
Here, the sum runs over all ions of charge $Z_i$ in the plasma, $g_e$
is the electron coupling parameter, which in ordinary cgs units (not
the rationalized units employed by BPS) is
\begin{equation}
  g_e = \frac{e^2 \, \kappa_e}{T_e} \,.
\end{equation}
The square of the thermal velocity entering Eq.~(\ref{classicdone})
is defined by 
\begin{eqnarray}
  V_{ei}^2 
  =
  \frac{T_e}{m_e} + \frac{T_\smI}{m_i} \ ,
\label{V}
\end{eqnarray}
while the quantity
\begin{eqnarray}
  \frac{1}{m_{ei}}
  =
  \frac{1}{m_e} + \frac{1}{m_i} 
\end{eqnarray}
is the reciprocal of the reduced mass of an electron-ion pair. 

The second term in Eq.~(\ref{Cabclqm}), the long-distance, dielectric 
term that accounts for collective effects in the plasma, is given by
\begin{eqnarray}
  {\cal C}^\smC_{e\smI,\smR} 
  =
  \frac{1}{2\pi} 
  \int_{-\infty}^{\infty}  dv \, 
  \frac{\rho_e(v)\,\rho_\smI(v)}{\rho_\text{total}(v)} \,
  \frac{i}{2 \pi}\, F_\text{total}(v) \, 
  \ln\!\left\{ \frac{F_\text{total}(v)}{\kappa_e^2}\right\} \,.
\label{nuion}
\end{eqnarray}
The individual spectral weights in this expression are defined by
\begin{eqnarray}
  \rho_b(u) 
  = 
  \kappa_b^2\,\sqrt{\frac{ m_b}{2\pi \, T_b } }\, u \,
  \exp\left\{- m_b\, u^2 / 2 T_b \right\} \ ,
\label{rhodef}
\end{eqnarray}
the sum over ion components is denoted by 
\begin{equation}
  \rho_\smI(u) =  {\sum}_i\, \rho_i(u) \ ,
\label{rhoI}
\end{equation}
and so the total spectral weight is given by
\begin{eqnarray}
  \rho_\text{total}(u) = \rho_e(u) + \rho_\smI(u) \,,
\end{eqnarray}
where $\rho_e(u)$ is the electron contribution.  The function
$F_\text{total}(v)$ is related to the classical, leading-order plasma
dielectric permittivity by
\begin{eqnarray}
  k^2 \epsilon(k,k v) = k^2 + F_\text{total}(v) \,,
\end{eqnarray}
and it can be written in terms of the dispersion relation
\begin{eqnarray}
  F_\text{total}(v) 
  = 
  \int_{-\infty}^\infty  du \, 
  \frac{\rho_\text{total}(u)}{u - v - i\,0^+} \,.
\label{Ftot}
\end{eqnarray}

As explained in BPS \cite{bps} in the work about their Eq.~(12.44), in
the limit of small electron mass the long-distance contribution
reduces, in leading order in $m_e$, to the simple form
\begin{eqnarray}
  {\cal C}^{\smC \,\, \smL}_{e\smI,\smR} 
  =
  \frac{\kappa_e^2}{2\pi}
  \left(\frac{ m_e}{2\pi \, T_e} \right)^{1/2} 
  \int_{-\infty}^{\infty}  dv \, v \,
  \frac{i}{2 \pi} \,
  \Big[ \kappa_e^2 + F_\smI(v) \Big] \, 
  \ln\!\left\{ 1 + \frac{F_\smI(v)}{\kappa_e^2}\right\} \,,
\label{nnuion0}
\end{eqnarray}
in which $F_\smI(v)$ differs form $F_\text{total}(v)$ by simply
replacing $\rho_\text{total}(u)$ in the dispersion
relation~(\ref{Ftot}) by $\rho_\smI(u)$. The integrand in the leading
term (\ref{nnuion0}) is analytic in the upper-half plane and hence,
noting the asymptotic behavior $F_\smI(v) \to - \omega_\smI^2 / v^2$
as $v \to \infty$, the integral can be evaluated by contour
integration with the result that
\begin{eqnarray}
  {\cal C}^{\smC \,\, \smL}_{e\smI,\smR} 
  =
  -\frac{1}{2}\,\frac{\kappa_e^2}{2\pi} \,
  \left( \frac{ m_e}{2\pi \, T_e} \right)^{1/2} \, \omega_\smI^2 \,.
\label{smart}
\end{eqnarray}
This is the result (12.44) of BPS after correcting for the trivial
transcription error noted above. We express the general long distance
contribution (\ref{nuion}) in the form
\begin{eqnarray}
  {\cal C}^\smC_{e\smI,\smR} 
  =
  \frac{\kappa_e^2}{2\pi} \,\left( \frac{ m_e}{2\pi \, T_e} \right)^{1/2} \, 
  \omega_\smI^2 \,\left\{ - \frac{1}{2} + \Delta  \right\} \,,
\label{D}
\end{eqnarray}
which, upon comparing Eq.~(\ref{nuion}) with Eqs.~(\ref{nnuion0}) and
(\ref{smart}), defines~\footnote{
  Note that the result of Ref.~\cite{KA} contains nothing in the
  nature of the finite electron mass, long-distance collective plasma
  contribution $\Delta$.
}
\begin{eqnarray}
  \Delta
  &=&
  \frac{\kappa_e^2}{\omega_\smI^2} 
  \int_{-\infty}^{\infty}  dv \, v \, \frac{i}{2 \pi} \, 
  \Bigg[  e^{-  m_e v^2 / 2 T_e } \, 
  \frac{\rho_\smI(v)}{\rho_\text{total}(v)} \,
  \frac{F_\text{total}(v)}{\kappa_e^2} \, 
  \ln\!\left\{ \frac{F_\text{total}(v)}{\kappa_e^2}\right\}
\nonumber\\
&& \qquad\qquad\qquad\qquad
  -\left[1 + \frac{F_\smI(v)}{\kappa_e^2} \right] \, 
  \ln\!\left\{ 1 + \frac{F_\smI(v)}{\kappa_e^2}\right\} \Bigg] \,.
\label{deltadef}
\end{eqnarray}
The correction $\Delta$ vanishes as $m_e \to 0$.  All the various
ratios that appear within the outer square brackets are dimensionless.
The prefactor $\kappa_e^2 / \omega_\smI^2 $ has the dimensions an
inverse velocity squared, which combines with the integration measure
$dv \, v$ to produce a dimensionless quantity.  Hence $\Delta$ is
dimensionless as it must be, and therefore it is a function only of
dimensionless quantities. One might expect that the only relevant
dimensionless parameter is the ratio of the squares of the thermal
velocity of the ions and electrons, $T_\smI m_e/T_e m_\smI$. However,
the Debye wavenumbers are important and they involve the temperature
and the density. Therefore, with the ion species at a common
temperature $T_\smI$, the most general set of dimensionless quantities
is the ion-electron temperature ratio $T_\smI/T_e$, the ion-electron
number density ratios $n_i/n_e$, the ion-electron mass ratios
$m_i/m_e$, and the dimensionless ionic charges $Z_i = e_i/e$. Hence,
\begin{equation}
  \Delta 
  = 
  \Delta \left( \frac{T_\smI}{T_e} \,,\, 
  \left\{\frac{n_i}{n_e} \right\} \,,\, 
  \left\{\frac{m_i}{m_e}\right\}  \,,\, 
  \left\{ Z_i \right\} \right) \,.
\end{equation}
In the case of a single ion species, as considered by Dimonte and
Daligault in Ref.~\cite{dd}, the correction $\Delta$ depends only upon
$T_\smI/T_e$, $m_\smI/m_e$, and $Z_\smI$.  We have examined analytic
approximations for $\Delta$, but they are long and cumbersome and do
not provide insight into its structure.  Hence in what follows, we
shall present graphs of $\Delta$ obtained numerically for various
parameters.

\begin{figure}
\includegraphics[scale=0.45]{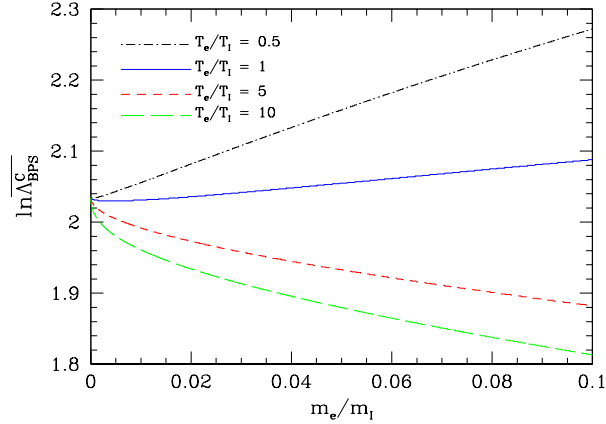}
\caption{\baselineskip 12pt plus 1pt minus 1pt 
  (Color)
  The BPS Coulomb logarithm $\ln \overline{\Lambda_\smBPS^\smC}$
  plotted as a function of $m_e/m_\smI$ for four values of the
  electron-ion temperature ratio, $T_e/T_\smI=0.5
  \,,\,1,5\,,\,10$, all with the coupling $g_e=0.1$. }
\label{fig:delta4}
\end{figure}
%%
% delta4.eps: delta4.sm, delta0.dat, delta1.dat, delta2.dat, 
%             delta3.dat, delta1.f90

\begin{figure}
\includegraphics[scale=0.45]{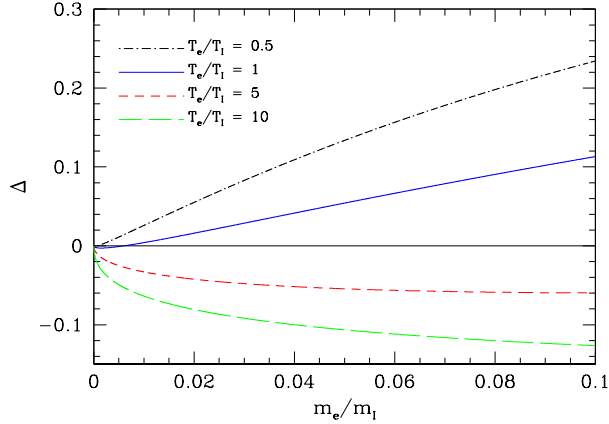}
\caption{\baselineskip 12pt plus 1pt minus 1pt 
  (Color)
  The correction $\Delta$ defined in Eq.~(\ref{deltadef}) plotted as a
  function of the mass ratio $m_e/m_\smI$ for a hydrogen plasma. The
  four curves correspond to the temperature ratios $T_e/T_\smI = 0.5, 
  1, 5, 10$.  }
\label{fig:delta}
\end{figure}

\begin{figure}
\includegraphics[scale=0.45]{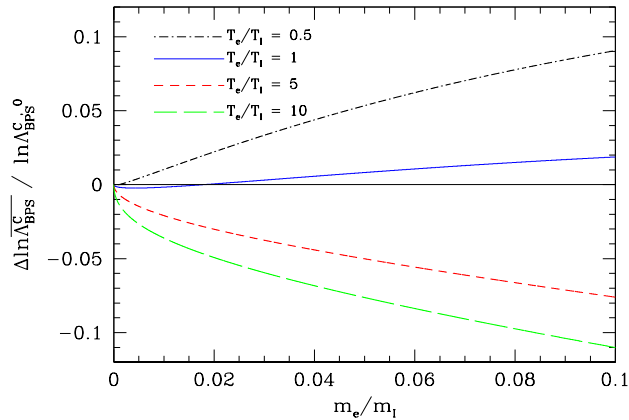}
\caption{\baselineskip 12pt plus 1pt minus 1pt 
  (Color)
  The relative electron-mass correction
  $\Delta\ln\overline{\Lambda_\smBPS^\smC}\,/ \ln \Lambda_\smBPS^{\smC
  \,,\,\smzero}$ plotted as a function of the mass ratio $m_e/m_\smI$
  for four values of the temperature ratio $T_e/T_\smI = 0.5, 1, 5,
  10$. In each case $g_e=0.1$. The mass correction
  $\Delta \ln\overline{ \Lambda_\smBPS^\smC}$ is defined in
  Eq.~(\ref{corrB}), while the zero-mass logarithm $\ln
  \Lambda_\smBPS^{\smC \,,\,\smzero}$ is given by Eq.~(\ref{corrA}).
  }
\label{fig:delta2}
\end{figure}

Adding Eq's.~(\ref{classicdone}) and (\ref{D}), and comparing with the
definition (\ref{rate}) of the Coulomb logarithm shows that in the
classical limit
\begin{eqnarray}
  \ln \Lambda^\smC_\smBPS
  &=& 
  {\sum}_i \, \frac{\omega_i^2}{\omega^2_\smI} \,
  \left(1 + \frac{T_\smI m_e}{T_e m_i}
  \right)^{-3/2} \,\left(\,\ln\!\left\{
  \frac{4 \, m_{ei} V^2_{ei}}{Z_i\,g_e \, T_e }\right\} - 2 \gamma\,  
  \right) - \frac{1}{2} +\Delta \,.
\label{LC}
\end{eqnarray}
This reduces to Eq.~(\ref{mezeroLC}) in the limit $m_e \to 0$.
Dimonte and Daligault\,\cite{dd} use the conventional definition of
the Coulomb logarithm for a single ion species rather than
convention~(\ref{rate}), one that applies to plasmas containing a
variety of ions. For a single species of ions, the two conventions are
related by
\begin{equation}
  \ln \Lambda^{\!\smC}
  = 
  \left(1 + \frac{T_\smI m_e}{T_e m_\smI} \,
  \right)^{-3/2} \, \ln \overline{\Lambda^{\!\smC}} \,,
\label{notationdiff}
\end{equation}
where we have denoted the conventional definition\,\footnote{
  As is apparent from Eq.~(\ref{LC}), the logarithm 
  depends upon ion species, and thus an overall factor of the form 
  $( 1 + T_\smI m_e/ T_e m_\smI)^{-3/2}$ 
  cannot be extracted in the general case.
} of the Coulomb logarithm by $\ln\overline{\Lambda^\smC}$.  Pulling
together previous definitions gives
\begin{eqnarray}
  \ln \overline{\Lambda^\smC_\smBPS}
  &=&  
  \ln\!\left\{ 
  \frac{4}{Z_\smI\,g_e } \right\} + \ln\!\left\{ \frac{m_\smI}{m_\smI + m_e} 
  \left(1 + \frac{T_\smI m_e}{T_e m_\smI} \right) \right\}
  - 2 \gamma +  \left(1 + \frac{T_\smI m_e}{T_e m_I}
  \right)^{3/2}\left[ - \frac{1}{2} +\Delta \right] \ ,
\nonumber\\
\label{firstcorr}
\end{eqnarray}
which in Fig.~\ref{fig:delta4} is plotted as a function of the
electron-ion mass ratio for several values of the temperature ratio.
Upon expanding to leading order in $m_e/m_\smI$ we can express the
Coulomb logarithm in terms of a zero electron-mass contribution and a
correction,
\begin{eqnarray}
  \ln \overline{\Lambda^\smC_\smBPS}
  &\simeq&
  \ln\Lambda^{\smC \,,\,\smzero}_\smBPS 
  +
  \Delta\ln\!\overline{\Lambda_\smBPS^\smC} \ ,
\label{leading}
\end{eqnarray}
where
\begin{eqnarray}
  \ln\Lambda^{\smC \,,\,\smzero}_\smBPS
  &=&
  \ln\!\left\{\frac{4}{Z_\smI\,g_e } \right\} - 2 \gamma -
  \frac{1}{2} 
\label{corrA}
\end{eqnarray}
is the zero electron-mass limit, and
\begin{eqnarray}
  \Delta\ln\overline{\Lambda_\smBPS^\smC}
  &=&  
  -\frac{m_e}{m_\smI}\left(1 - \frac{T_\smI}{4 T_e} \right) + \Delta 
\label{corrB}
\end{eqnarray}
is the leading order electron mass correction. 
Dimonte and Daligault \cite{dd} use
$g_e=0.1$ and consider the cases in which $T_e / T_\smI$ varies from
about 1 to 5 with $m_e / m_\smI $ varying from about zero to 0.02,
while $Z_\smI = 1 = n_e / n_\smI$ are fixed.  Figure~\ref{fig:delta}
displays the values of $\Delta$ about this parameter range.
%%
% delta1.eps: delta1.sm, delta0.dat, delta1.dat, delta2.dat, 
%             delta3.dat, delta1.f90
%
Figure~\ref{fig:delta2} presents the complete leading corrections for
the Coulomb logarithm (\ref{firstcorr}) as the mass ratio $ m_e /
m_\smI $ is varied --- Eq.~(\ref{corrB}) divided by
Eq.~(\ref{corrA}). The leading term (\ref{corrA}) is about 2.0, and so
the relative correction is correspondingly smaller.
Figure~\ref{fig:delta2} shows that the relative size of the electron
mass correction in the range examined by Dimonte and Daligault
\cite{dd} is less than 2\%, which is less than their statistical
accuracy of 5\%. With smaller statistical error, one could resolve
the mass effects (\ref{corrB}) with an MD simulation.
%%
% delta2.eps: delta2.sm, delta0.dat, delta1.dat, delta2.dat, 
%             delta3.dat, delta1.f90

\begin{acknowledgments}
  We would like to thank D. Preston for a number of 
  useful conversations. We would also like to thank
  G. Dimonte, J. Daligault, and J. Reynold for comments
  on the manuscript.
\end{acknowledgments}

\pagebreak[4]
\appendix

\section{Results Collected}

We put together here all the formulae relevant to the cases of
physical interest in which the scattering is predominantly quantum
mechanical.  These include the small classical corrections to this
purely quantum limit and the leading effects of Fermi-Dirac statistics
that come into play as the electron density is increased. This we do
because, with the inclusion of the $\Delta$ correction, we now have in
hand all the small corrections to the leading quantum-mechanical
scattering limit.  To exhibit these, we write
\begin{equation}
  \ln\Lambda_{\smBPS} 
  = 
  \ln\Lambda^\smQM_{\smBPS} + \ln\Lambda^{\smDelC}_{\smBPS} + 
  \ln\Lambda^{\smFD}_{\smBPS} \,.
\label{qresult}
\end{equation}
Here $\ln\Lambda^\smQM_{\smBPS}$ is the leading term in the quantum
limit together with the $\Delta$ correction that we have exhibited in
the text, $\ln\Lambda^{\smDelC}_{\smBPS}$ is the first classical
correction that appears when the parameters depart from the extreme
quantum limit, and $\ln\Lambda^{\smFD}_{\smBPS}$ is the first
correction when Fermi-Dirac statistics start to become important. The
latter two terms have been computed in Ref.~\cite{bs} to leading order in
the small ratio $m_e / m_i $; this suffices since the terms are
already themselves small.

In the text we examined the limit of purely classical scattering and
thus omitted the quantum correction term ${\cal C}^{\smDelQ}_{e \,
\smI}$ in Eq.~(\ref{Cabclqm}) from the complete relaxation rate. As a
first step in presenting the collection mentioned, we quote this
omitted correction which is Eq.~(12.50) of BPS:
\begin{eqnarray}
  {\cal C}^\smDelQ_{e \,\smI}
  &\!\!=\!&
  \!- {\sum}_i \frac{\kappa_e^2\, \kappa_i^2}{2} \, 
  \frac{(T_\smI^2 m_e \, T_e^2 m_i)^{1/2}}{(T_\smI m_e \!+\! 
  T_e m_i)^{3/2}}\! 
  \left(\frac{1}{2\pi}\right)^{\!\!3/2} \!\!
  \int_0^\infty \!\!\! d \zeta\, e^{-\zeta/2} \left[{\rm Re}\,\, 
  \psi\!\left\{\!1 + i\frac{\bar\eta_{ei}}{\zeta^{1/2}}\right\} \!-\! 
  \ln\!\left\{ \frac{\bar\eta_{ei}}{\zeta^{1/2}}\right\} \right] .
\nonumber\\
\label{qqcorr}
\end{eqnarray}
Here $\psi(z) = d \ln \Gamma(z) / dz$, and 
\begin{eqnarray}
  \bar\eta_{ei} = \frac{|e e_i|}{\hbar V_{ei}}  
\end{eqnarray}
makes precise the definition of the quantum parameter alluded to at
the beginning of the text with the square of the thermal velocity in
this expression $V_{ei}^2$ defined previously in Eq.~(\ref{V}).  The
extreme quantum limit in which $\bar\eta_{ei} \to 0$ of this formula
is spelled out in detail in Ref.~\cite{bps}. Here we shall not repeat
the derivation but simply quote the BPS limit (12.53) with slightly
different notation:
\begin{eqnarray}
  {\cal C}^\smDelQ_{e \, \smI}
  &=&  
  {\sum}_i \frac{\kappa_e^2\, \kappa_i^2}{2} \, 
  \frac{(T_\smI^2 m_e \, T_e^2 m_i)^{1/2}}{(T_\smI m_e \!+\! T_e m_i)^{3/2}}\! 
  \left(\frac{1}{2\pi}\right)^{3/2} \, \left[ 3 \gamma + 
  \ln\!\left\{ \frac{Z_i^2 e^4 }{2 \hbar^2 \, V_{ei}^2} \right\} \right]  \,.
\label{qqqcorr}
\end{eqnarray}
This quantum correction, added to the classical scattering
contribution (\ref{classicdone}), produces
\begin{equation}
  {\cal C}^\smQ_{e\smI,\smS} 
  = 
  {\sum}_i \frac{\kappa_e^2 \, \kappa_i^2}{2} \, 
  \frac{ (T_\smI^2 m_e T_e^2 m_i)^{1/2}}{\left( T_\smI m_e + 
  T_e m_i \right)^{3/2} } \,\left( \frac{1}{2\pi} \right)^{3/2}\, 
  \left[\ln\!\left\{\left(\frac{ 8\, T_e^2}{\hbar^2 \omega_e^2}\right)
  \left(\frac{m^2_{ei}}{m_e} \frac{V^2_{ei}}{T_e} \right) \right\} - 
  \gamma\,  \right] \,. 
\end{equation}
In the same way that the classical Coulomb logarithm (\ref{LC}) was
constructed, the quantum scattering version now reads
\begin{eqnarray}
  \ln \Lambda^\smQM_\smBPS
  =
  {\sum}_i \, \frac{\omega_i^2}{2 \omega_\smI^2} \,
  \left(1 + \frac{T_\smI m_e}{T_e m_i}
  \right)^{-3/2} \,\left[\ln\!\left\{\left(
  \frac{ 8\, T_e^2}{\hbar^2 \omega_e^2}\right)
  \left(\frac{m^2_{ei}}{m_e} \frac{V^2_{ei}}{T_e} \right) \right\}  
  - \gamma\,  \right] - \frac{1}{2} + \Delta \,.
\end{eqnarray}
The explicit electron-ions mass ratio terms that appear here
(including those contained in the definition of $m_{ei}$ and
$V^2_{ei}$\,) are easy to compute.  For typical ICF conditions, they
make very small corrections on the order or less than 1\%. So as to
make the significance of the $\Delta$ correction clear, a correction
that does require some computation, we now omit these small explicit
terms and write
\begin{eqnarray}
  \ln \Lambda^\smQM_\smBPS
  =
  \frac{1}{2} \, \left[\ln\!\left\{  \frac{ 8\, T_e^2}{\hbar^2 \omega_e^2}
  \right\}   - \gamma -1 \,  \right]  + \Delta \,,
\label{done}
\end{eqnarray}
which is precisely the result (\ref{bpsrateQM}) of the text, but with
the additional finite electron mass correction $\Delta$.

\begin{figure}
\includegraphics[scale=0.45]{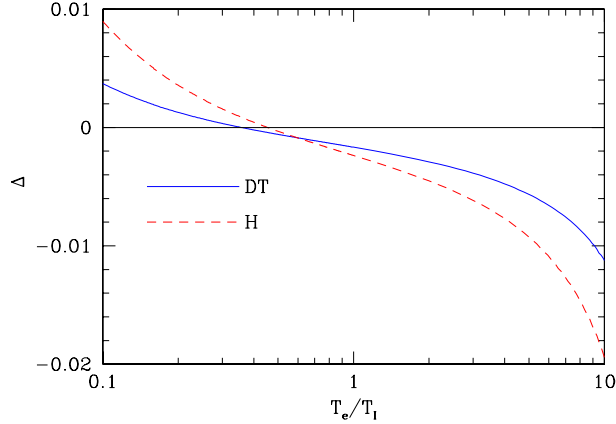}
\caption{\baselineskip 12pt plus 1pt minus 1pt 
  (Color)
  The correction $\Delta$ defined in Eq.~(\ref{deltadef}) for an
  equimolar DT plasma (solid) and a hydrogen plasma (dashed) plotted
  as a function $T_e/T_\smI$ for the physical values of the electron
  and ion masses.  }
\label{fig:delta3}
\end{figure}

We show the $\Delta$ correction in Fig.~\ref{fig:delta3} over a wide
range of the temperature ratio $T_\smI/T_e$ for the typical ICF case
of an equimolar DT plasma. For a burning plasma, the Coulomb logarithm
has the rough value $ \ln \Lambda^\smQM_\smBPS \approx 4 $, and so the
relative $\Delta$ correction is about a quarter of the number shown in
Fig.~\ref{fig:delta3}.
%%
% delta3.eps: delta3.sm, delta4.dat, delta5.dat, delta2.f90

For the remaining terms in Eq.~(\ref{qresult}), we shall just quote
the result from Eq.~(2.6) presented in Ref.~\cite{bs}, namely
\begin{equation}   
  \ln\Lambda^{\smDelC}_{\smBPS} = -\,\frac{\epsilon_\smH}{T_e}\, {\sum}_i \, 
  \frac{Z_i^2 \, \omega_i^2}{\omega_\smI^2} \,
  \left[ \zeta(3)\left(
  \ln\!\left\{\frac{T_e}{Z_i^2\,\epsilon_\smH} \right\}
  -\gamma \right) -2\,\zeta'(3) \right] \,,
\label{cc}
\end{equation}
and~\footnote{
  Reference \cite{bs} also contains the result in which the
  quantum-mechanical scattering is computed with exact Fermi-Dirac
  statistics, not just the first correction away from
  Maxwell-Boltzmann statistics which we quote here.
}
\begin{equation}
  \ln\Lambda^{\smFD}_{\smBPS} =
  \frac{n_e \lambda_e^3}{2} \, \left[-\left( 1 - \frac{1}{2^{3/2}} \right)\,
  \frac{1}{2}\left[ \ln\!\left\{  \frac{8 T_e^2}{\hbar^2\omega_e^2}\right\}
  -\gamma - 1  \right] + \left(\frac{1}{2}\,\ln 2  + 
  \frac{1}{2^{5/2}} \right) \right] \,.
\label{ss}
\end{equation}
The ratio $\epsilon_\smH/T_e$ describes the relative size of the first
quantum to classical correction, where
\begin{eqnarray}
  \epsilon_\smH 
  = 
  \,\frac{e^4 \, m_e}{2 \, \hbar^2} 
  \simeq 13.6 \, {\rm eV}
\label{bind}
\end{eqnarray}
is the binding energy of the hydrogen atom.  The numerical values of
the zeta-function and its derivative are
\begin{eqnarray}
  \zeta(3) 
  =  
  \sum_{k=1}^\infty \, \frac{1}{k^3}  = 1.20205 \cdots \,,
\end{eqnarray}
and
\begin{eqnarray}
  \zeta'(3) 
  =  
  -\sum_{k=1}^\infty \, \frac{1}{k^3} \, \ln k = -0.19812 \cdots \,.
\end{eqnarray}
The electron thermal wave length
\begin{equation}
\lambda_e = \hbar \left(\frac{2\pi}{m_e T_e}\right)^{1/2} 
\end{equation}
sets the scale at which quantum statistics comes into play, with $z_e
= n_e \lambda_e^3 / 2$ the electron fugacity.

For some temperature and number density regimes of interest, the two
corrections (\ref{cc}) and (\ref{ss}) become comparable in
size\,\cite{bs}.

\end{document}